\documentstyle[12pt]{article}
\begin{document}

\baselineskip=24pt plus 2pt
\hfill\hbox{NCKU-HEP/97-05}
\begin{center}

{\large \bf On the energy of the de Sitter-Schwarzschild black hole}\\
\vspace{5mm}
\vspace{5mm}
I-Ching Yang \footnote{E-mail:icyang@ibm65.phys.ncku.edu.tw}

Department of Physics, National Cheng Kung University \\
Tainan, Taiwan 701, Republic of China \\

\end{center}
\vspace{5mm}

\begin{center}
{\bf ABSTRACT}
\end{center}
   Using Einstein's and Weinberg's energy complex, we evaluate the 
energy distribution of the vaccum nonsingularity black hole solution.
The energy distribution is positive everywhere and be equal to zero at
origin.

\vspace{2mm}
\newpage

   There are two reasons for evaluting the energy of a system in
general relativity. First, the conserved qualitities, like total energy 
etc., play important roles in solving the equation of motion. Second,
the energy distribution must be positive if attractive force only.
In particular positions, $ r = 2M $ and $ r = 0 $, the Schwarzschild
solution will be degenerate. The value $ r = 2M $ is a removable
coordinate singularity, however, the singularity at the origin is
indeed irremoved. All of the physical laws are not holden at the 
singularity, hence we can't predict the evolution of singularity.
The energy distribution at the singularity can't be prediction,
therefore, the gravity theory without singularity is inevitable.
Many theories, which are like Brans-Dicke theory, Einstein-Cartan
theory etc., try to avoid the existence of the singularity, but they
can not do it.
With considering the action in which pure gravity term adds to 
cosmological constant term, the first vacuum nonsingularity solution
is written down by de Sitter~\cite{1} in 1917. The de Sitter geometry 
is generated by a vacuum with nonzero energy density
$ \varepsilon = \frac{\Lambda}{8\pi} $, 
described by the stress-energy tensor
\begin{equation}
  T_{\alpha\beta} = \varepsilon g_{\alpha\beta} ,
\end{equation}
with the equation of state
\begin{equation}
  p = -\varepsilon .
\end{equation} 

   Recent, Dymnikova~\cite{1} shows that the spherically symmetric 
vacuum can generate a black hole solution which is regular at $ r = 0 $ 
and everywhere else. In the spherically symmetric static case, a line 
element has the form
\begin{equation}
  ds^2 = e^\nu dt^2 - e^\lambda dr^2 - r^2 d\theta^2 -r^2 sin\theta d\phi^2 ,
\end{equation}
with the boundary conditions are the Schwarzschild behavior at 
$ r \rightarrow \infty $,
\begin{eqnarray}
  ds^2 = (1 - \frac{r_g}{r})dt^2 -\frac{dr^2}{1 - \frac{r_g}{r}} - r^2 d\theta^2 - r^2 sin\theta d\phi^2 , \\
  T_{\mu\nu} = 0,
\end{eqnarray}
where $ r_g = 2M $, and the de Sitter behavior at $ r \rightarrow 0 $
\begin{eqnarray}
  ds^2 = ( 1 - \frac{r^2}{r_0^2})dt^2 - \frac{dr^2}{1 - \frac{r^2}{r_0^2}} - r^2 d\theta^2 - r^2 sin\theta d\phi^2 , \\
  T_{\mu\nu} = \varepsilon_0 g_{\mu\nu}.
\end{eqnarray}
The limiting density is fixed and connected with the de Sitter horizon
parameter $ r_0 $ by the de Sitter relation
\begin{equation}
  r_0^2 = \frac{3}{8\pi\varepsilon_0} .
\end{equation}
With the assumed form of the energy-momentum tensor
\begin{equation}
  T^0_0 = \varepsilon_0 exp(- \frac{r^3}{r_0^2 r_g}) ,
\end{equation}
they obtain the following metric
\begin{equation}
  ds^2 = ( 1 - \frac{R_g (r)}{r} ) dt^2 - \frac{1}{ 1 - \frac{R_g (r)}{r}} dr^2 - r^2 d\theta^2 - r^2 sin\theta d\phi^2 .
\end{equation}
In this solution, the defination of the parameters are 
\begin{equation}
  R_g (r) = r_g ( 1 - exp( \frac{r^3}{r_*^3} )) 
\end{equation}
and
\begin{equation}
  r_*^3 = r_0^2 r_g .
\end{equation}
The structure of this solution, the so-called de Sitter-Schwarzschile
 solution, is like
a Schwarzschild solution whose singularity is replaced by the de Sitter
core. In this literary, we study the energy distribution of the de 
Sitter-Schwarzschild solution by the energy-momentum pseudotensor. 

   We consider the energy componens according to the energy-momentum
pseudotensors of Einstein~\cite{2}, 
\begin{equation}
  E(r) = \frac{1}{16\pi} \int \frac{\partial H_0^{0l}}{\partial x^l} d^3x ,
\end{equation}
where 
\begin{equation}
  H_0^{0l} = \frac{g_{00}}{\sqrt{-g}} \frac{\partial}{\partial x^m} \left[ (-g) g^{00} g^{lm} \right] ,
\end{equation}
and the energy-momentum pseudotensors of Weinberg~\cite{3},
\begin{equation}
  E(r) = - \frac{1}{8\pi} \int \frac{\partial Q_0^{0l}}{\partial x^l} d^3x ,
\end{equation}
where
\begin{eqnarray}
  Q_0^{0l} = \frac{1}{2} ( h_{lm,m} - h_{mm,l} ) , \\
  h_{lm} = g_{lm} - \eta_{lm} .
\end{eqnarray}
The Latin index takes valuse from 1 to 3. 
We carry out the energy components according to energy-momentum
pseudotensors of Einstein and Weinberg calculation in the 
quasi-Cartesian coordinate $ (t,x,y,z) $. The line element (10)
converted into quasi-Cartesian coordinate is
\begin{equation}
  ds^2 = (1 - \frac{R_g (r)}{r}) dt^2 - (dx^2 + dy^2 + dz^2) - \left( \frac{R_g (r)}{r^2 (r - R_g (r))} \right) (xdx + ydy + zdz)^2 .
\end{equation}
Thus, we obtain the required nonvanishing components of Einstein's
energy-momentum pseudotensor $ H_0^{0l} $ in Eq. (14)
\begin{eqnarray}
  H_0^{01} = \frac{2x}{r^3} R_g (r) , \\
  H_0^{02} = \frac{2y}{r^3} R_g (r) , \\
  H_0^{03} = \frac{2z}{r^3} R_g (r) .   
\end{eqnarray}
Weinberg's energy-momentum pseudotensor $ Q_0^{0l} $ in Eqs. (16)-(17)
\begin{eqnarray}
  Q_0^{01} = - \frac{x}{r^3} R_g (r) , \\
  Q_0^{02} = - \frac{y}{r^3} R_g (r) , \\
  Q_0^{03} = - \frac{z}{r^3} R_g (r) .
\end{eqnarray}
After plugging the nonvanishing components of energy-momentum
pseudotensors into the formulation of energy complex, and applying the  
Gauss theorem, we evaluate the integral over the surface of a sphere
with radius $ r $. 

    Finally, we obtain the energy complex defined by Einstein and
Weinberg within a sphere with radius $ r $ are the same, and the result
is  
\begin{equation}
  E(r) = \frac{r_g}{2} \left( 1 - exp(-\frac{r^3}{r_*^3}) \right) ,
\end{equation}
and this result is the same as the standard formula for the mass~\cite{4}
\begin{eqnarray}
  m(r) & = & 4\pi \int_0^r T_0^0 r^2 dr  \\
       & = & \frac{r_g}{2} \left( 1 - exp(- \frac{r^3}{r_*^3}) \right).
\end{eqnarray}
   We plot the energy distributions of de Sitter-Schwarzschild black
hole, see Figure . The behavior of energy distribution is like the 
Schwarzschild solution 
\begin{equation}
  E(r) = M
\end{equation}
at $ r \rightarrow \infty $, and the de Sitter solution (we show the
energy distribution of de Sitter solition in the appendix of this 
literary.) 
\begin{equation}
  E(r) = \frac{r^3}{2r_0^2}
\end{equation}
at $ r \rightarrow 0 $. By Einstein's and Weinberg's defination,
we find that the energy distribution is positive everywhere, even in
the region $ r < r_H $, and be equal to zero at origin. This solution
shows only attractive force everywhere including the origin in pure
gravity theory.

\begin{center}
{\bf Acknowledgements}
\end{center}
I thanks Prof. R.R. Hsu for useful comments and discussions.

\newpage
\begin{center}
{\bf Appendix}
\end{center}
   The de Sitter solution is the following metric 
\begin{equation}
  ds^2 = (1-\frac{r^2}{r_0^2})dt^2 - \frac{dr^2}{1-\frac{r^2}{r_0^2}} - r^2 d\theta^2 - r^2 sin \theta d\phi^2
\end{equation}
in spherical coordinate, and 
\begin{equation}
  ds^2 = (1-\frac{r^2}{r_0^2})dt^2 - (dx^2+dy^2+dz^2) - (\frac{1}{r_0^2-r^2})(xdx+ydy+zdz)^2
\end{equation}
in quasi-Cartesian coordinate, where $ r_0^2 = \frac{3}{\Lambda} $ and
$ \Lambda $ is the cosmological constant.
According Eq.(14), we can obtain the 
nonvanishing components of Einstein's energy-momentum pseudotensor
$ H_0^{0l} $
\begin{eqnarray}
  H_0^{01} = \frac{2x}{r_0^2} , \\
  H_0^{02} = \frac{2y}{r_0^2} , \\
  H_0^{03} = \frac{2z}{r_0^2} .
\end{eqnarray}
Plugging those nonvanishing components into Eq.(13), and applying
the Gauss theorem, we evaluate the integral over the surface of a 
sphere within radius $ r $. The energy complex defined by Einstein 
within a sphere with radius $ r $ is 
\begin{equation}
  E(r) = \frac{2r^3}{r_0^2} .
\end{equation}

\newpage
\begin{figure}[hp]
\setlength{\unitlength}{0.240900pt}
\ifx\plotpoint\undefined\newsavebox{\plotpoint}\fi
\sbox{\plotpoint}{\rule[-0.200pt]{0.400pt}{0.400pt}}%
\begin{picture}(1500,900)(0,0)
\font\gnuplot=cmr10 at 10pt
\gnuplot
\sbox{\plotpoint}{\rule[-0.200pt]{0.400pt}{0.400pt}}%
\put(220.0,113.0){\rule[-0.200pt]{292.934pt}{0.400pt}}
\put(220.0,113.0){\rule[-0.200pt]{0.400pt}{184.048pt}}
\put(220.0,113.0){\rule[-0.200pt]{4.818pt}{0.400pt}}
\put(198,113){\makebox(0,0)[r]{0}}
\put(1416.0,113.0){\rule[-0.200pt]{4.818pt}{0.400pt}}
\put(220.0,215.0){\rule[-0.200pt]{4.818pt}{0.400pt}}
\put(198,215){\makebox(0,0)[r]{0.2}}
\put(1416.0,215.0){\rule[-0.200pt]{4.818pt}{0.400pt}}
\put(220.0,317.0){\rule[-0.200pt]{4.818pt}{0.400pt}}
\put(198,317){\makebox(0,0)[r]{0.4}}
\put(1416.0,317.0){\rule[-0.200pt]{4.818pt}{0.400pt}}
\put(220.0,419.0){\rule[-0.200pt]{4.818pt}{0.400pt}}
\put(198,419){\makebox(0,0)[r]{0.6}}
\put(1416.0,419.0){\rule[-0.200pt]{4.818pt}{0.400pt}}
\put(220.0,520.0){\rule[-0.200pt]{4.818pt}{0.400pt}}
\put(198,520){\makebox(0,0)[r]{0.8}}
\put(1416.0,520.0){\rule[-0.200pt]{4.818pt}{0.400pt}}
\put(220.0,622.0){\rule[-0.200pt]{4.818pt}{0.400pt}}
\put(198,622){\makebox(0,0)[r]{1}}
\put(1416.0,622.0){\rule[-0.200pt]{4.818pt}{0.400pt}}
\put(220.0,724.0){\rule[-0.200pt]{4.818pt}{0.400pt}}
\put(198,724){\makebox(0,0)[r]{1.2}}
\put(1416.0,724.0){\rule[-0.200pt]{4.818pt}{0.400pt}}
\put(220.0,826.0){\rule[-0.200pt]{4.818pt}{0.400pt}}
\put(198,826){\makebox(0,0)[r]{1.4}}
\put(1416.0,826.0){\rule[-0.200pt]{4.818pt}{0.400pt}}
\put(220.0,113.0){\rule[-0.200pt]{0.400pt}{4.818pt}}
\put(220,68){\makebox(0,0){0}}
\put(220.0,857.0){\rule[-0.200pt]{0.400pt}{4.818pt}}
\put(463.0,113.0){\rule[-0.200pt]{0.400pt}{4.818pt}}
\put(463,68){\makebox(0,0){1}}
\put(463.0,857.0){\rule[-0.200pt]{0.400pt}{4.818pt}}
\put(706.0,113.0){\rule[-0.200pt]{0.400pt}{4.818pt}}
\put(706,68){\makebox(0,0){2}}
\put(706.0,857.0){\rule[-0.200pt]{0.400pt}{4.818pt}}
\put(950.0,113.0){\rule[-0.200pt]{0.400pt}{4.818pt}}
\put(950,68){\makebox(0,0){3}}
\put(950.0,857.0){\rule[-0.200pt]{0.400pt}{4.818pt}}
\put(1193.0,113.0){\rule[-0.200pt]{0.400pt}{4.818pt}}
\put(1193,68){\makebox(0,0){4}}
\put(1193.0,857.0){\rule[-0.200pt]{0.400pt}{4.818pt}}
\put(1436.0,113.0){\rule[-0.200pt]{0.400pt}{4.818pt}}
\put(1436,68){\makebox(0,0){5}}
\put(1436.0,857.0){\rule[-0.200pt]{0.400pt}{4.818pt}}
\put(220.0,113.0){\rule[-0.200pt]{292.934pt}{0.400pt}}
\put(1436.0,113.0){\rule[-0.200pt]{0.400pt}{184.048pt}}
\put(220.0,877.0){\rule[-0.200pt]{292.934pt}{0.400pt}}
\put(45,495){\makebox(0,0){E(r)}}
\put(828,23){\makebox(0,0){r}}
\put(220.0,113.0){\rule[-0.200pt]{0.400pt}{184.048pt}}
\put(220,113){\usebox{\plotpoint}}
\put(232,112.67){\rule{3.132pt}{0.400pt}}
\multiput(232.00,112.17)(6.500,1.000){2}{\rule{1.566pt}{0.400pt}}
\put(245,113.67){\rule{2.891pt}{0.400pt}}
\multiput(245.00,113.17)(6.000,1.000){2}{\rule{1.445pt}{0.400pt}}
\put(257,115.17){\rule{2.500pt}{0.400pt}}
\multiput(257.00,114.17)(6.811,2.000){2}{\rule{1.250pt}{0.400pt}}
\multiput(269.00,117.60)(1.651,0.468){5}{\rule{1.300pt}{0.113pt}}
\multiput(269.00,116.17)(9.302,4.000){2}{\rule{0.650pt}{0.400pt}}
\multiput(281.00,121.59)(1.123,0.482){9}{\rule{0.967pt}{0.116pt}}
\multiput(281.00,120.17)(10.994,6.000){2}{\rule{0.483pt}{0.400pt}}
\multiput(294.00,127.59)(0.758,0.488){13}{\rule{0.700pt}{0.117pt}}
\multiput(294.00,126.17)(10.547,8.000){2}{\rule{0.350pt}{0.400pt}}
\multiput(306.00,135.58)(0.543,0.492){19}{\rule{0.536pt}{0.118pt}}
\multiput(306.00,134.17)(10.887,11.000){2}{\rule{0.268pt}{0.400pt}}
\multiput(318.00,146.58)(0.497,0.493){23}{\rule{0.500pt}{0.119pt}}
\multiput(318.00,145.17)(11.962,13.000){2}{\rule{0.250pt}{0.400pt}}
\multiput(331.58,159.00)(0.492,0.669){21}{\rule{0.119pt}{0.633pt}}
\multiput(330.17,159.00)(12.000,14.685){2}{\rule{0.400pt}{0.317pt}}
\multiput(343.58,175.00)(0.492,0.755){21}{\rule{0.119pt}{0.700pt}}
\multiput(342.17,175.00)(12.000,16.547){2}{\rule{0.400pt}{0.350pt}}
\multiput(355.58,193.00)(0.492,0.927){21}{\rule{0.119pt}{0.833pt}}
\multiput(354.17,193.00)(12.000,20.270){2}{\rule{0.400pt}{0.417pt}}
\multiput(367.58,215.00)(0.493,0.933){23}{\rule{0.119pt}{0.838pt}}
\multiput(366.17,215.00)(13.000,22.260){2}{\rule{0.400pt}{0.419pt}}
\multiput(380.58,239.00)(0.492,1.099){21}{\rule{0.119pt}{0.967pt}}
\multiput(379.17,239.00)(12.000,23.994){2}{\rule{0.400pt}{0.483pt}}
\multiput(392.58,265.00)(0.492,1.186){21}{\rule{0.119pt}{1.033pt}}
\multiput(391.17,265.00)(12.000,25.855){2}{\rule{0.400pt}{0.517pt}}
\multiput(404.58,293.00)(0.493,1.131){23}{\rule{0.119pt}{0.992pt}}
\multiput(403.17,293.00)(13.000,26.940){2}{\rule{0.400pt}{0.496pt}}
\multiput(417.58,322.00)(0.492,1.272){21}{\rule{0.119pt}{1.100pt}}
\multiput(416.17,322.00)(12.000,27.717){2}{\rule{0.400pt}{0.550pt}}
\multiput(429.58,352.00)(0.492,1.272){21}{\rule{0.119pt}{1.100pt}}
\multiput(428.17,352.00)(12.000,27.717){2}{\rule{0.400pt}{0.550pt}}
\multiput(441.58,382.00)(0.492,1.272){21}{\rule{0.119pt}{1.100pt}}
\multiput(440.17,382.00)(12.000,27.717){2}{\rule{0.400pt}{0.550pt}}
\multiput(453.58,412.00)(0.493,1.131){23}{\rule{0.119pt}{0.992pt}}
\multiput(452.17,412.00)(13.000,26.940){2}{\rule{0.400pt}{0.496pt}}
\multiput(466.58,441.00)(0.492,1.142){21}{\rule{0.119pt}{1.000pt}}
\multiput(465.17,441.00)(12.000,24.924){2}{\rule{0.400pt}{0.500pt}}
\multiput(478.58,468.00)(0.492,1.056){21}{\rule{0.119pt}{0.933pt}}
\multiput(477.17,468.00)(12.000,23.063){2}{\rule{0.400pt}{0.467pt}}
\multiput(490.58,493.00)(0.493,0.893){23}{\rule{0.119pt}{0.808pt}}
\multiput(489.17,493.00)(13.000,21.324){2}{\rule{0.400pt}{0.404pt}}
\multiput(503.58,516.00)(0.492,0.884){21}{\rule{0.119pt}{0.800pt}}
\multiput(502.17,516.00)(12.000,19.340){2}{\rule{0.400pt}{0.400pt}}
\multiput(515.58,537.00)(0.492,0.712){21}{\rule{0.119pt}{0.667pt}}
\multiput(514.17,537.00)(12.000,15.616){2}{\rule{0.400pt}{0.333pt}}
\multiput(527.58,554.00)(0.492,0.625){21}{\rule{0.119pt}{0.600pt}}
\multiput(526.17,554.00)(12.000,13.755){2}{\rule{0.400pt}{0.300pt}}
\multiput(539.00,569.58)(0.497,0.493){23}{\rule{0.500pt}{0.119pt}}
\multiput(539.00,568.17)(11.962,13.000){2}{\rule{0.250pt}{0.400pt}}
\multiput(552.00,582.58)(0.600,0.491){17}{\rule{0.580pt}{0.118pt}}
\multiput(552.00,581.17)(10.796,10.000){2}{\rule{0.290pt}{0.400pt}}
\multiput(564.00,592.59)(0.758,0.488){13}{\rule{0.700pt}{0.117pt}}
\multiput(564.00,591.17)(10.547,8.000){2}{\rule{0.350pt}{0.400pt}}
\multiput(576.00,600.59)(0.874,0.485){11}{\rule{0.786pt}{0.117pt}}
\multiput(576.00,599.17)(10.369,7.000){2}{\rule{0.393pt}{0.400pt}}
\multiput(588.00,607.60)(1.797,0.468){5}{\rule{1.400pt}{0.113pt}}
\multiput(588.00,606.17)(10.094,4.000){2}{\rule{0.700pt}{0.400pt}}
\multiput(601.00,611.60)(1.651,0.468){5}{\rule{1.300pt}{0.113pt}}
\multiput(601.00,610.17)(9.302,4.000){2}{\rule{0.650pt}{0.400pt}}
\put(613,615.17){\rule{2.500pt}{0.400pt}}
\multiput(613.00,614.17)(6.811,2.000){2}{\rule{1.250pt}{0.400pt}}
\put(625,617.17){\rule{2.700pt}{0.400pt}}
\multiput(625.00,616.17)(7.396,2.000){2}{\rule{1.350pt}{0.400pt}}
\put(638,618.67){\rule{2.891pt}{0.400pt}}
\multiput(638.00,618.17)(6.000,1.000){2}{\rule{1.445pt}{0.400pt}}
\put(650,619.67){\rule{2.891pt}{0.400pt}}
\multiput(650.00,619.17)(6.000,1.000){2}{\rule{1.445pt}{0.400pt}}
\put(662,620.67){\rule{2.891pt}{0.400pt}}
\multiput(662.00,620.17)(6.000,1.000){2}{\rule{1.445pt}{0.400pt}}
\put(220.0,113.0){\rule[-0.200pt]{2.891pt}{0.400pt}}
\put(674.0,622.0){\rule[-0.200pt]{183.566pt}{0.400pt}}
\end{picture}
 \caption{Thr energy distribution of de Sitter-Schwarzschild solution with $ r_g = 2 $ and $ r_0 = \frac{1}{2} $.}
\end{figure}
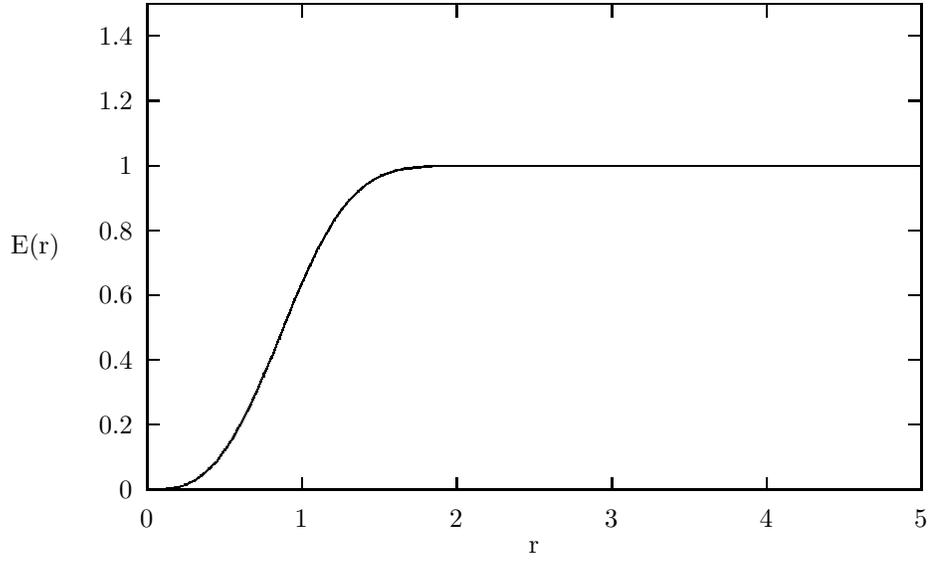


\begin{thebibliography}{set}
\bibitem{1}
I.G. Dymnikova, {\it Gen. Rel. Grav.} {\bf 24}, 235(1992).
\bibitem{2}
C. M{\o}ller, {\it Ann. Phys.}(NY) {\bf 4}, 347(1958).
\bibitem{3}
S. Weinberg, {\it Gravitation and Cosmology: Principles and Applications
 of General Theory of Relativity}(John Wiley and Sons, Inc., NY, 1972).
\bibitem{4}
L.D. Landau and E.M. Lifshitz, {\it The Classical Theory of Fields}
 (Pergamon Press, Oxford, 1975).
\end{thebibliography}
\end{document}